\documentclass[11pt]{article}
\usepackage{moriond,epsfig}

\def\be{\begin{equation}}
\def\ee{\end{equation}}
\def\bea{\begin{eqnarray}}
\def\eea{\end{eqnarray}}

\def\LEPI{LEP$\!I$}
\def\LEPII{LEP$\!I\!I$}
\newcommand{\leptlept}{\ensuremath{\ell^+\ell^-}}
\newcommand{\epem}{\ensuremath{\mathrm{e}^+\mathrm{e}^-}}

\newcommand{\bb}{\ensuremath{b\bar{b}}}
\newcommand{\cc}{\ensuremath{c\bar{c}}}
\newcommand{\qq}{\ensuremath{q\bar{q}}}

\newcommand{\eecc}{\epem\rightarrow\cc}
\newcommand{\eebb}{\epem\rightarrow\bb}
\newcommand{\eegg}{\epem\rightarrow \gamma\gamma}
\newcommand{\eeee}{\epem\rightarrow\epem}

\newcommand{\eeqq}{\epem\rightarrow\qq}
\newcommand{\eell}{\epem\rightarrow \leptlept}

\newcommand{\jth}{j_{\rm had}^{\rm tot}}

\begin{document}
\vspace*{4cm}
\title{STANDARD MODEL AT \boldmath \LEPII \unboldmath}

\author{ K. Sachs }

\address{Department of Physics, Carleton University, \\
1125 Colonel By Drive, Ottawa, Ontario K1S 5B6, Canada}

\maketitle\abstracts{
High statistics Standard Model processes like fermion- and photon-pair 
production in $\epem$ collisions are studied at centre-of-mass energies 
up to 209 GeV. No significant deviation from the Standard Model is 
observed, leading to strong constraints on models with additional 
heavy  vector bosons, contact interactions and
low scale gravity. A largely model independent determination of the 
Z mass becomes feasible with the measurement of $\gamma / Z$ 
interference in $\rm\epem\to\bar{q}q$ at \LEPII . 
}

\section{Introduction}

The large electron positron collider LEP was build to study
properties of the Z with very high precision. This was achieved 
with the \LEPI\ program in the years 1989 to 1995 at centre-of-mass
energies around the Z mass. During the years 1996 to 2000, referred to
as \LEPII\ period, the energy was increased up to a maximum of 209 GeV
yielding high sensitivity both for direct and indirect searches
for physics beyond the Standard Model. A total luminosity 
of almost 700 pb$^{-1}$ was recorded by four experiments at energies 
above the W-pair threshold.

High precision measurements of cross-sections for fermion- and
photon-pair final states are used to search for signals of new physics.
Since the systematic errors are only partially correlated a combination
of the statistically independent results of the four experiments leads
to a significant improvement of the sensitivity. In combination with
the high accuracy of the Standard Model prediction this leads to 
a search range well beyond the LEP energy region.

\section{Fermion-pair final states}

\begin{table}[t]
\caption{Expected final experimental precision and theoretical
uncertainty for LEP combined results of different channels of
fermion-pair production. The numbers for electron-pair
production are for the region $|\cos{\theta}|<-0.7$.
\label{tab:error}}
\vspace{0.4cm}
\begin{center}
\begin{tabular}{|l|cccc|cc|}
\hline 
 & $\rm\sigma(q)$ & $\rm\sigma(e)$ & $\rm\sigma(\mu)$ &  
   $\rm\sigma(\tau)$ & $A_{\rm fb}(\mu)$ & $A_{\rm fb}(\tau)$ 
\\\hline
experiment
 & 1\% & 0.9\% & 1.6\% & 2.2\% & 0.012 & 0.015 \\
theory
 & 0.26\% & 2.0\% & 0.4\% & 0.4\% & 0.004 & 0.004 
\\\hline
\end{tabular}
\end{center}
\end{table}

\begin{figure}[b]
\epsfxsize0.45\textwidth\epsffile{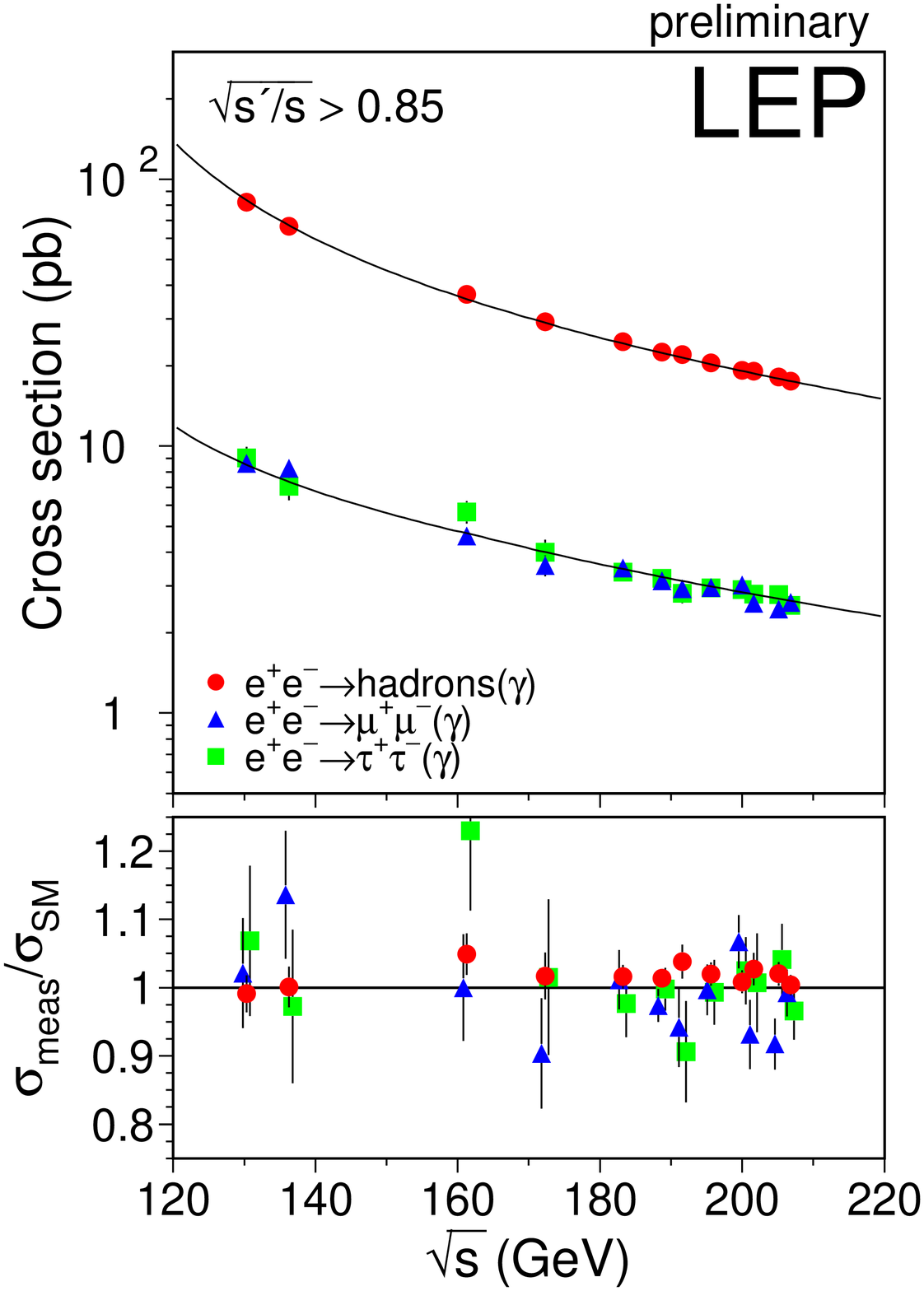}\hfill
\epsfxsize0.45\textwidth\epsffile{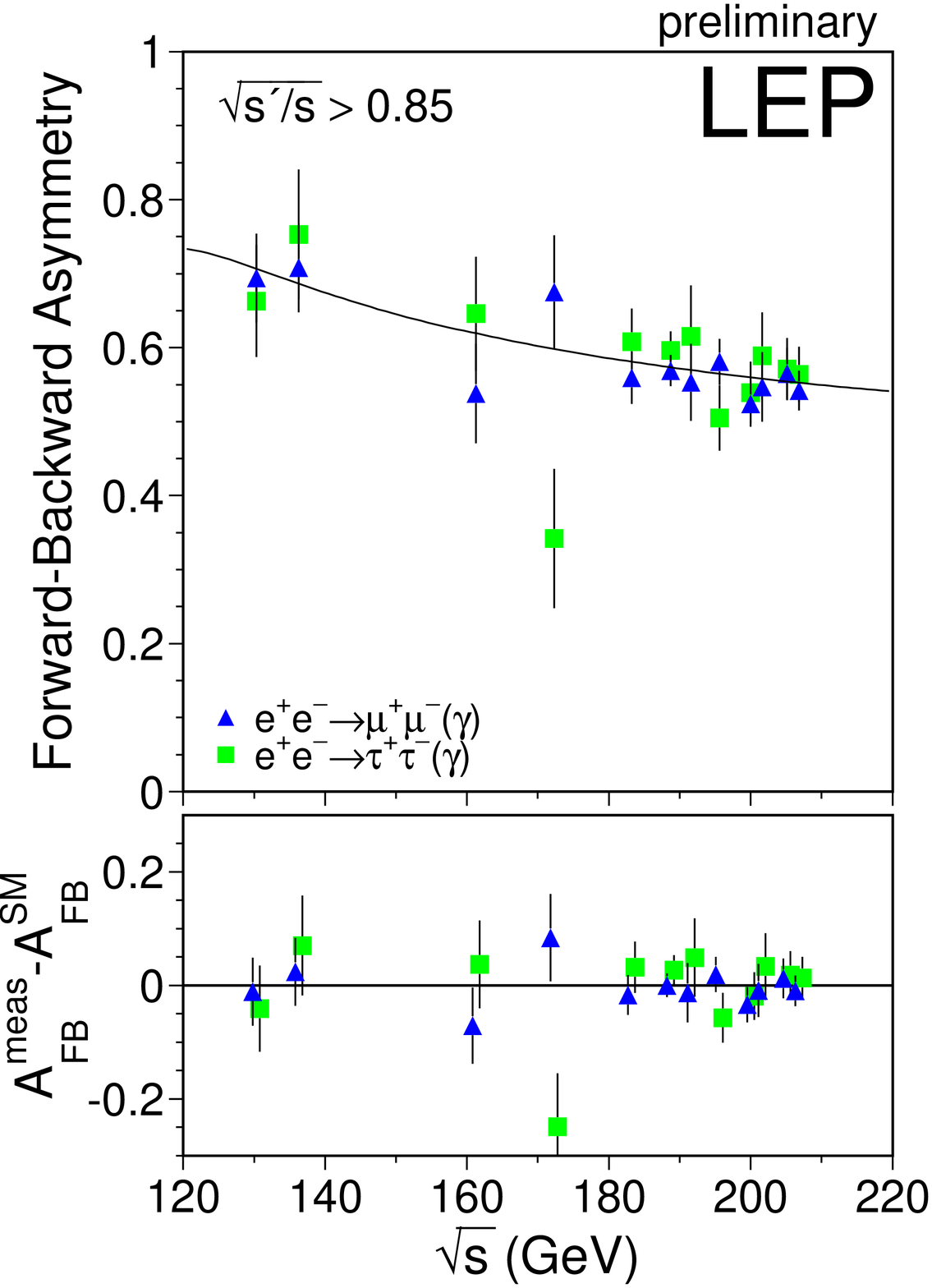}
\caption{LEP combined total cross-sections and forward-backward
asymmetries for high $s'$ hadronic, $\mu$- and $\tau$-pair 
final states at energies between 130 and 209 GeV.
\label{fig:xsnafb}}
\end{figure}

Fermion-pair production at \LEPI\ was clearly dominated by
the production and decay of an on-shell Z boson. At higher 
energies the process becomes more complicated due to strong 
initial state radiation, ISR, final state radiation, FSR, and a larger 
contribution of photon exchange leading to stronger interference 
effects. Initial state radiation results in a reduction of the effective
centre-of-mass energy, $\sqrt{s'}$. For $\sqrt{s'} \sim M_{\rm Z}$
the cross-section increases, leading to a radiative-return peak.
Since the coupling of quarks to the Z is much stronger than to the
photon, compared to charged leptons, the radiative return is more
pronounced for hadronic final states than for $\mu$- and $\tau$- pairs.
Only about 22\% of the hadronic events have high $s'/s > 0.85$; for
leptons this fraction is about 41\%. The Bhabha cross-section is dominated
by $t$-channel exchange and hence doesn't show a pronounced radiative
return peak. To enhance the sensitivity to new physics only high $s'$
events are discussed in the following report.
Events from neutrino-pair production with ISR are mainly studied
in the context of direct searches for invisible particles and are
not covered here.
An additional effect is initial-state pair-production, ISPP,
leading to an effective four-fermion final state. However, ISPP should
be treated similar to ISR as long as the propagator is a photon, not a 
Z boson. 
Careful treatment of these effects, both in the theoretical
calculations and in the experimental measurement, lead to an 
improvement of the errors on the cross-section predictions. 
The expected experimental precision of the final results of the 
LEP combination are compared to the current theoretical errors \cite{ref:yelrep} 
in Table \ref{tab:error}. For all processes, apart from Bhabha 
scattering, the theory error is well below the experimental precision.
For endcap Bhabhas the theory error is smaller, around 0.5\%, and
the experimental precision is limited by systematic errors inspite of
high statistics.

LEP combined results for the differential cross-sections of
Bhabhas, $\mu$- and $\tau$-pairs at energies between 183 and 209 GeV
show good agreement with the Monte Carlo expectation \cite{ref:ff}.
Measured total cross-sections and forward-backward asymmetries are
shown in Figure \ref{fig:xsnafb} and compared to the prediction from
ZFITTER \cite{ref:zfitter}. Similar results are available for heavy
flavour final states, 
$R_{\rm c}$, $R_{\rm b}$, $A^{\rm c}_{\rm fb}$ and $A^{\rm b}_{\rm fb}$,
though not all inputs are provided by all collaborations \cite{ref:ff}.

Results of cross-sections and asymmetries are used to search for
indirect signals of physics beyond the Standard Model which would
in most cases arise from a new particle as a propagator. Examples 
are leptoquarks being exchanged in the $t$- or $u$-channel in the
production of hadronic events, Kaluza-Klein gravitons in models with
large scale extra dimensions or an additional heavy vector boson Z$'$.
A general approach to test such models is the study of four-fermion
contact interactions with the effective Lagrangian:
$${\cal L}_{\rm eff} = \frac{g^2}{(1+\delta)\Lambda^2}
\sum_{i,j=L,R} \eta_{ij}\bar{e}_i\gamma_\mu e_i\bar{f}_j\gamma^\mu f_j
\; ,$$
where the coupling is fixed to $g=4\pi$, $\delta=1$ for Bhabhas and 0
else, $\eta_{ij} = 0, \pm 1$ describe the chiral structure and $\Lambda$ is the
scale of the interaction. The limits summarized in Table~\ref{tab:ci} 
range from about 2 to 20~TeV.

\begin{table}[t]
\caption{Limits on contact interaction for different final states:
Bhabhas, $\mu$ and $\tau$ pairs (labeled as $\ell^+\ell^-$),
hadrons, c and b quarks. 
Models are specified by the combination of 
$(\eta_{\rm LL},\eta_{\rm RR},\eta_{\rm LR},\eta_{\rm RL})$:
\mbox{LL = $(\pm 1,0,0,0)$}, \mbox{RR = $(0,\pm 1,0,0)$},
\mbox{LR = $(0,0,\pm 1,0)$}, \mbox{RL = $(0,0,0,\pm 1)$},
\mbox{VV = $(\pm 1,\pm 1,\pm 1,\pm 1)$}, 
\mbox{AA = $(\pm 1,\pm 1,\mp 1,\mp 1)$}, 
\mbox{V0 = $(\pm 1,\pm 1,0,0)$} and \mbox{A0 = $(0,0,\pm 1,\pm 1)$}.
\label{tab:ci}}
\vspace{0.4cm}
\begin{center}
\setlength{\tabcolsep}{2ex}
\begin{tabular}{|c||cc|cc|cc|cc|cc|}
\hline
 (TeV) & \multicolumn{2}{c||}{$\eeee$} &
        \multicolumn{2}{c||}{$\eell$} &
        \multicolumn{2}{c||}{$\eeqq$} &
        \multicolumn{2}{c||}{$\eecc$} &
        \multicolumn{2}{c|}{$\eebb$} \\
\hline
  Model   & $\Lambda^{-}$ & $\Lambda^{+}$ 
          & $\Lambda^{-}$ & $\Lambda^{+}$ 
          & $\Lambda^{-}$ & $\Lambda^{+}$ 
          & $\Lambda^{-}$ & $\Lambda^{+}$ 
          & $\Lambda^{-}$ & $\Lambda^{+}$ \\
\hline 
\hline    
~~~~LL~~~~ &  9.0  & 7.1  
           &  9.8  & 13.3 
           & 3.7  & 6.0  
           &    5.7 &    6.6 
           &    9.1 &   12.3 \\
\hline
    RR     & 8.9  & 7.0  
           &  9.3  & 12.7 
           & 5.5  & 3.9  
           &    4.9 &    1.5 
           &   2.2 &    8.1 \\
\hline                                                     
    VV     &18.0  &15.9  
           & 16.0  & 21.7  
           & 8.1  & 5.3  
           &    8.2 &   10.3 
           &    9.4 &   14.1 \\
\hline
    AA     &11.5  &11.3  
           & 15.1  & 17.2  
           & 5.1  & 8.8  
           &    6.9 &    7.6 
           &  11.5 &   15.3 \\
\hline                                                     
    LR     &10.0  & 9.1  
           &  8.6  & 10.2 
           & 5.1  & 4.3  
           &   3.9 &    2.1 
           &   3.1 &    5.5 \\
\hline
    RL     & 10.0  & 9.1  
           &   8.6  & 10.2 
           &  7.2  & 9.3  
           &    3.1 &    2.8 
           &    7.0 &    2.4 \\
\hline                                                     
    V0     & 12.5  &10.2  
           &  13.5  & 18.4  
           &  5.1  & 6.0  
           &     7.4 &    9.2 
           &   10.8 &   14.5 \\
\hline
    A0     & 14.0  &13.0  
           &  12.4  & 14.3
           &  8.0  & 3.9  
           &    4.5 &    2.7 
           &   6.3 &    3.9 \\
\hline
\end{tabular}
\end{center}
\end{table}

Limits are also derived within the specific models.
The couplings of leptoquarks can be restricted depending on the
leptoquark mass. Bhabha scattering gives the best sensitivity for
models with large extra dimensions, restricting the energy scale to
1.2 TeV and 1.09 TeV depending on the model. Cross-sections and
asymmetries measured at \LEPII\ are sensitive to the interference
of a Z$'$ with the ordinary Z leading to limits on the Z$'$ mass 
between 342 and 1787 GeV, depending on the model. In general a 
Z$'$ can mix with the Standard Model Z, thus changing the
lineshape and couplings observed at \LEPI . A lineshape fit to
\LEPI\ data within the Z$'$ model provides information about this mixing 
angle $\theta_{\rm M}$. Since no combined cross-sections and asymmetries 
for \LEPI\ data are available this study is performed by separate 
experiments only. Limits are in the range $|\theta_{\rm M}| <4$ mrad
\cite{ref:delphi,ref:l3,ref:opal}.

The most precise determination of the Mass of the Z is derived from 
the \LEPI\ lineshape fit to hadronic and leptonic cross-sections 
and leptonic asymmetries. Most of the \LEPI\ data were recorded in 
three centre-of-mass energy regions:
at the peak and about 1.8 GeV above and below. Since the coupling 
of the Z is stronger to quarks than to charged leptons about 88\% 
of all data are hadronic events at these three energies, resulting
in effectively three measurements dominating the determination of the 
lineshape parameters which describe the mass of the Z, $M_{\rm Z}$, 
its width $\Gamma_{\rm Z}$, its coupling to quarks via the total 
hadronic cross-section $\sigma^0_{\rm had}$ and leptonic parameters.
The mentioned three parameters can be accurately determined from
the three measurements if all other aspects are assumed to be known.
This includes the couplings of the photon to quarks and the interference
between photon and Z contributing to the hadronic cross-section.

\begin{figure}[bh]
\centerline{\mbox{\epsfxsize0.6\textwidth\epsffile{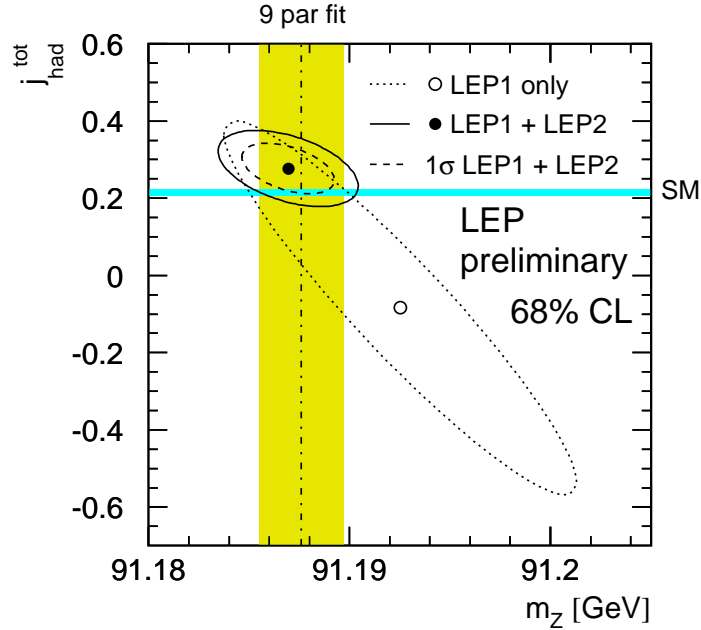}}}
\caption{Correlation between the mass of the Z and 
$j^{\rm tot}_{\rm had}$. Results are shown for \LEPI\
data only and for a combined fit to \LEPI\ and \LEPII\ data. 
The yellow band indicates the 1 $\sigma$ error from the 9 parameter fit.
\label{fig:smat}}
\end{figure}

From the results of the standard lineshape fit it is difficult to
determine which effect a change in the interference or additional
contributions from physics beyond the Standard Model might have.
Therefore a more model independent ansatz, the S-matrix formalism
\cite{ref:smatasy}, 
is applied to determine the Z mass. It assumes two neutral bosons as
propagators, one of them massless, but leaves the contribution of
the Z exchange, $r_{\rm had}^{\rm tot}$, and the
interference, $\jth$, to the hadronic cross-section explicitly free.
However, four parameters cannot be accurately determined from three
effective measurements, leading to a large correlation between
$M_{\rm Z}$ and $\jth$ shown in Figure~\ref{fig:smat}. Including
more measurements from \LEPII\ solves this problem, reducing the
correlation. The final result of 
$M_{\rm Z} = 91\;186.9 \pm 2.3 \mbox{ MeV}$ \cite{ref:smat} is in 
very good agreement with the result of the standard lineshape fit
$M_{\rm Z} = 91\;187.6 \pm 2.1 \mbox{ MeV}$ \cite{ref:9par}
with only slightly increased error.

\section{Photon-pair final states}
The process $\eegg$ is purely electromagnetic with weak contributions 
at the considered energies of about 0.2\%. The theory is well known,
the coupling constant $\alpha$ at zero momentum 
transfer \footnote{Photons don't appear in the propagator but as real
particles in the final state.} is precisely measured. However,
cross-section predictions are available only for Born level and next 
to leading order \cite{ref:berends}. The theoretical error is proportional
to the applied radiative corrections, which depend on the selected 
phase space, especially the acolinearity restriction. 
Table \ref{tab:gg_phase}
summarizes the phase space selected by the four experiments.
Though the corrections from the next to leading order prediction
to Born level are different for the experiments, the preliminary
combination assumes a common theory error of 1\%.

\begin{table}[t]
\caption{Selected phase space of the four experiments depending on the 
photon angle $\theta$, the acolinearity $\xi$ and the photon energies
$E$. OPAL replaces the cut on the acolinearity by a restriction on the
missing longitudinal momentum. The resulting
radiative corrections in the barrel ($|\cos{\theta}| < 0.6$) range 
from an up correction of 5\% (ALEPH)
to a down correction of 28\% (L3).\label{tab:gg_phase}}
\begin{center}
\renewcommand{\arraystretch}{1.2}
\begin{tabular}{|c|cccc|}
\hline
Experiment & polar angle & acolinearity & energy & rad. correction
\\ \hline \hline
ALEPH & $|\cos{\theta}| < 0.95$ & $\xi < 20^\circ$ & 
$E > 0.25\sqrt{s}$ & -5\% to -2\% \\ \hline
DELPHI & \raisebox{0ex}[4ex][2.5ex]
{\parbox{4cm}{$0.035 < |\cos{\theta}| < 0.731$\\
$0.819 < |\cos{\theta}| < 0.906$}} & $\xi < 50^\circ$ & 
$E > 0.15\sqrt{s}$ & 5\% to 8\% \\ \hline
L3    & $|\cos{\theta}| < 0.961$ & $\xi < 175^\circ$ & 
\raisebox{0ex}[4ex][2.5ex]
{\parbox{2.8cm}{$E > 5\mbox{ GeV}$ \\ $\Sigma E > 0.5\sqrt{s}$}} &
18\% to 28\% \\ \hline
OPAL  & $|\cos{\theta}| < 0.93$ & $p_{\rm l} < E_{1,2}$ & 
\raisebox{0ex}[4ex][2.5ex]
{\parbox{2.8cm}{$E > 1\mbox{ GeV}$ \\ $\Sigma E + p_{\rm l}>
0.6\sqrt{s}$}}&
4\% to 7\% \\ \hline
\end{tabular}
\end{center}
\end{table}

\begin{figure}[bp]
\centerline{\mbox{\epsfxsize0.6\textwidth\epsffile{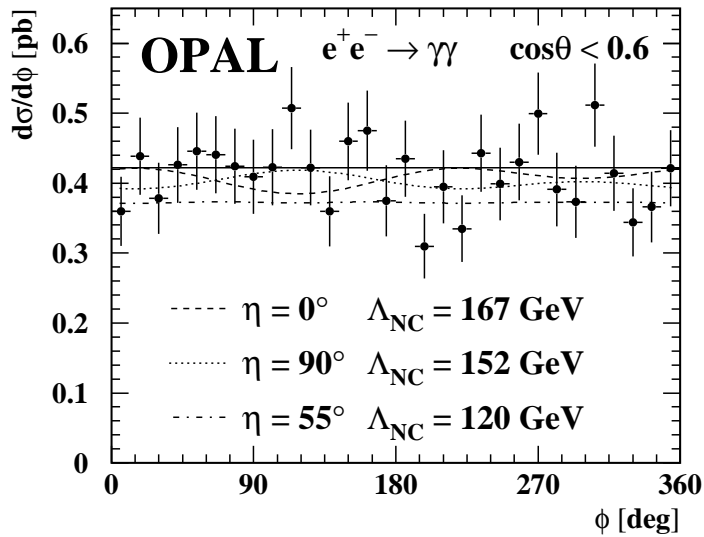}}}
\caption{Dependence of the $\eegg$ cross-section on $\phi$
in comparison to different expectations of non-commutative QED.
\label{fig:ncqed}}
\end{figure}

The combination \cite{ref:gg}
based on final results from ALEPH,
L3 and OPAL as well as preliminary 
results from DELPHI yields an average of 
0.982$\pm$0.010 for the ratio of the measured over the expected total
cross-section at energies between 183 and 209 GeV.
The differential cross-sections determined by the four experiments
are analyzed in a combined fit to set limits on scales of 
physics beyond the Standard Model: 
anomalous couplings $\Lambda_+ >392 \mbox{ GeV}$,
$\Lambda_- >364 \mbox{ GeV}$, 
contact interactions $\Lambda'>831 \mbox{ GeV}$, 
low scale gravity $M_{\rm s}>933 \mbox{ GeV} (\lambda=+1)$, 
$M_{\rm s}>1010 \mbox{ GeV} (\lambda=-1)$, 
and the coupling $f/\Lambda < 3.9 \mbox{ TeV}^{-1}$ 
of excited electrons with a mass $M_{\rm e^\ast} = 200 \mbox{ GeV}$.
The best limits on $f/\Lambda$ are derived from LEP direct searches
but are restricted to the kinematic region 
$M_{\rm e^\ast} < 210 \mbox{ GeV}$ while the indirect limits from
$\eegg$ depend only weakly on $M_{\rm e^\ast}$.

All of these models affect only $d \sigma / d \cos{\theta}$.
However, there are also models like non-commutative QED, which
lead to deviations of the azimuthal angle $\phi$.
Non-commutative geometry is predicted by the quantization of strings 
in the presence of a background field \cite{ref:ncqed_string}. 
This background field
defines a unique direction which leads to a dependence of the
cross-section not only on $\theta$ but also on $\phi$, on time
and the orientation of the detector. A non-commutative
Standard Model has not yet been formulated, but QED based on a
non-commutative geometry (NCQED) exists, albeit with some limitations:
only integer charges are allowed and the size of higher order
corrections is unclear. 

Figure \ref{fig:ncqed} shows the cross-section as a function of $\phi$
integrated over $|\cos{\theta}|<0.6$ and averaged over time
\cite{ref:ncqed_O}. It is compared to lowest order predictions of 
NCQED \cite{ref:ncqed_LO} for different
angles $\eta$ between the unique direction and the rotation axis of
the earth. For OPAL the $\cos{\theta}$ distribution provides
a $\eta$ independent limit on the scale $\Lambda > 141 \mbox{ GeV}$ which
can be improved by the $\phi$ distribution up to  
$\Lambda > \mbox{167 GeV}$ for $\eta = 0^\circ$.

\section{Conclusion}
LEP data taking was finalized in 2000 and the final \LEPII\ results 
now start to come up. Besides the study of W-pair production and
direct searches the precise measurements of Standard Model processes
like fermion- and photon-pair production contributed to the large
variety of processes studied. \LEPII\ is a prime example
of how four collaborations can act as in fact one experiment.
The careful combination of results in the LEPEW working group
exploits fully the power of electroweak measurements possible under
the clean conditions of an $\epem$ collider. As a result the
high precision of the data forced the theorists to improve the
Standard Model predictions to the sub \% level.
Fermion- and photon-pair cross-sections mainly provide indirect
limits which complement the direct searches done at the Tevatron and
HERA. No discovery was made at LEP, but we gained a much better
understanding of the Standard Model.


\end{document}